\documentclass[preprint,12pt]{elsarticle}
\usepackage{amsmath}
\usepackage{amsthm}
\usepackage{tabularx}
\usepackage{hyperref}
\hypersetup{
    colorlinks=true,
    linkcolor=blue,
    filecolor=magenta,      
    urlcolor=cyan,
    pdftitle={MuMuPy: A Dimuonium-Matter Interaction Calculator},
    pdfpagemode=FullScreen,
    }

\usepackage{amssymb}
\usepackage{listings}
\usepackage{xcolor}

\definecolor{codegreen}{rgb}{0,0.6,0}
\definecolor{codegray}{rgb}{0.5,0.5,0.5}
\definecolor{codepurple}{rgb}{0.58,0,0.82}
\definecolor{backcolour}{rgb}{0.94,0.95,0.94}

\lstdefinestyle{mystyle}{
    backgroundcolor=\color{backcolour},   
    commentstyle=\color{codegreen},
    keywordstyle=\color{magenta},
    numberstyle=\tiny\color{codegray},
    stringstyle=\color{codepurple},
    basicstyle=\ttfamily\footnotesize,
    breakatwhitespace=false,         
    breaklines=true,                 
    captionpos=b,                    
    keepspaces=true,                 
    numbers=left,                    
    numbersep=5pt,                  
    showspaces=false,                
    showstringspaces=false,
    showtabs=false,                  
    tabsize=2
}

\usepackage{amssymb}

\newcounter{bla}

\journal{Computer Physics Communications}

\begin{document}

\begin{frontmatter}

\title{MuMuPy: A Dimuonium-Matter Interaction Calculator}

\author[a,b]{Artem Uskov\corref{author}}
\author[a]{Abdaljalel Alizzi}
\author[a,b]{Zurab Silagadze}

\cortext[author] {Corresponding author.\\\textit{E-mail address:} a.uskov+articles@alumni.nsu.ru}
\address[a]{Novosibirsk State University, Novosibirsk 630 090, Russia}
\address[b]{Budker Institute of Nuclear Physics, Budker Institute of Nuclear Physics}

\begin{abstract}
In this article, we present MuMuPy, a computational library and cloud-based tool for calculating cross sections for the interaction of dimuonium (true muonium) with matter. MuMuPy calculates corresponding form factors and allows one to find the probabilities of dimuonium transitions in the electric field of the nucleus. 

MuMuPy was developed in the context of the $\mu\mu$-tron facility, the project of a low-energy electron-positron collider for production and experimental study of dimuonium, proposed in our home institute, Budker Institute of Nuclear Physics.

The reliability of MuMuPy was verified by three independent methods, one of which was developed by the authors earlier.
\end{abstract}


\end{frontmatter}

{\bf PROGRAM SUMMARY}

\begin{small}
\noindent
{\em Program title:} MuMuPy \\ 
{\em CPC Library link to program files:}                             \\
{\em Code Ocean capsule:}  https://codeocean.com/capsule/1734849/tree/v1    \\ 
{\em Licensing provisions:} GPLv3 \\
{\em Programming language:}  Python 3.X, Fortran  \\
{\em Nature of problem:} Planned experiments at the $\mu\mu$-tron facility include investigating the interaction of dimuonium with ordinary atoms as it passes through the foil. This requires the creation of a computational infrastructure for a reliable calculation of the interaction of dimuonium with the electric field of the atomic nucleus. \\  
{\em Solution method:} Three independently developed computational strategies were used, including our own new method based on the identities of hypergeometric functions. A MuMuPy library and a cloud-based tool for calculating the cross sections for the interaction of dimuonium with matter have been developed, which we hope will be used in future studies of true muonium. 
\end{small}

\section{Introduction}
\label{intro}
Dimuonium, an as yet undiscovered exotic atom, is an interesting object for modern physics, especially in the light of the signs of yet unexplained anomalies in the muon sector \cite{abi_measurement_2021,aebischer_b-decay_2020,baum_tiny_2021,aebischer_effective_2021}. In this respect, upcoming true muonium experiments \cite{vidal_discovering_2019} are in line with plans to impose additional constraints and pinpoint new physics beyond the Standard Model  \cite{kanda_new_2021,delaunay_towards_2021}.

Our home institute, Budker Institute of Nuclear Physics, has an ambitious plan to build $\mu\mu$-tron, a low-energy cross-angle electron-positron collider for the discovery and research of dimuonium \cite{bogomyagkov_low-energy_nodate}. Using state-of-the art accelerator technologies such as Crab-Waist with a large crossing angle, it becomes possible to detect ($\mu^+\mu^-$)-bound state, providing sufficient luminosity and dramatically reducing the background.

This exotic elementary atom could be a gift to the high energy physics community that is 
trying to tame observed unexplained anomalies in the muon sector by building Strandard Model extensions and/or theorizing new particles. The leptonic anomalies mentioned include:
the measured anomalous magnetic moment of the muon for a long time contradicts the predictions of the Standard Model theory \cite{abi_measurement_2021, Aoyama:2020ynm}, a large number of rare B-meson decays via $b \to s \mu^+\mu^- $ transitions  apparently deviate from the predictions of the Standard Model \cite{LHCb:2021trn,Altmannshofer:2021qrr} and may indicate a violation of lepton-flavor universality, the proton radius puzzle (a disagreement of experimental values of  proton charge radius measured in normal and muonic hydrogen) \cite{Gao:2021sml,Khabarova}. This circumstance adds motivation to the  upcoming experiments with true muonium. We hope that these experiments will enable us to impose more restrictions on new physics in the muon sector.

One of the planned experiments at $\mu\mu$-tron is the interaction of dimuonium with matter as it passes through the foil. This experiment will allow several valuable properties of dimuonium to be investigated. For the software support of this experiment, we will need a thoroughly tested set of tools to calculate the appropriate interaction cross sections. As a result, we have developed the MuMuPy library. 

The calculation of the form factors of hydrogen-like atoms is well studied in physics. In MuMuPy, we use not only well-known algorithms, but also a new algorithm proposed by the authors earlier. As a result, a fully functional package was created for calculating the cross sections for the interaction of dimuonium with the Coulomb field of a target nucleus in a non-relativistic approximation, the key feature of which is the reliability provided by three independent implementations of the computational algorithms and the corresponding subsequent tests. 

In the following sections of this article, we will briefly explain why research on dimuonium is relevant to physics, describe our new computation method and our approach to creating a cloud interface for the MuMuPy library with examples. 

\section{Dimuonium - the smallest QED atom}
Dimuonium, bimuonium or true muonium is a electromagnitically bound muon-antimuon atom
\cite{Baier,Malenfant:1987tm,Brodsky:2009gx}. Why is this exotic atom so special?
\begin{enumerate}[1.]
\item It has never been observed. There are six elementary leptonic atoms in total: $e^+e^-$, $\mu^+e^-$, $\mu^+ \mu^-$, $\tau^+ e^-$, $\tau^+ \mu^-$ and $\tau^+ \tau^-$. Only positronium $e^+e^-$ and muonium $\mu^+e^-$ were discovered.

\item There are no strongly interacting particles in it. Dimuonium is truly a quantum electrodynamical system. Quantum chromodynamics manifests itself only in the form of corrections through the relatively well-studied hadronic vacuum polarization. Electroweak corrections to the hyperfine splitting of true muonium may have experimental significance, but they can be calculated \cite{Lamm:2015lia}. This circumstance makes dimuonium more theoretically computable from the point of view of transition rates, production and decay cross sections \cite{lamm_true_2015}. 

\item  Studying dimuonium could be an excellent precision test of QED, and not only. 

For example, one of the precision QED tests is the measurement of hyperfine splitting as a result of the magnetic interaction between the total angular momentum of an electron and the nuclear spin.  In hydrogen, the measurement of hyperfine splitting makes it possible to determine the value of the fine structure constant as  $\alpha^{-1} = 137.0360 (3)$.
Note that the internal structure of the proton makes theoretical calculations uncertain, which leads to a relatively large experimental uncertainty in the determination of $\alpha$. 

Unlike the hydrogen atom, whose nucleus is composed of quarks, muonium consists only of point-like particles. This significantly increases the accuracy of measuring the fine structure constant from the data of hyperfine splitting of muonium: $\alpha^{-1} = 137.035 994 (18)$.

\item 
Dimuonium studies can  provide an independent verification of the muon anomalous magnetic moment results. Currently, the muon (g-2) discrepancy reaches $4.2\sigma$, and nears an exciting $5\sigma$ discovery level \cite{abi_measurement_2021}. To check this result, it would be desirable to use another experimental method with an independent set of systematic errors. 

As is known, hyperfine splitting is due to the interaction of the magnetic moments $\mu_1$ and $\mu_2$ of the components of the bound system (see the interaction Hamiltonian
(\ref{H}), $\mu_0$ is the vacuum permeability). 
\begin{equation}
{H}_{HFS} = -\frac{\mu_{0}}{4 \pi} \frac{1}{r^{3}}\left[3\left(\vec{\mu}_{1} \cdot \vec{r}\right)\left(\vec{\mu}_{2} \cdot \vec{r}\right)-\vec{\mu}_{1} \cdot \vec{\mu}_{2}\right]-\frac{2}{3} \, \mu_{0}\, \vec{\mu}_{1} \cdot \vec{\mu}_{2} \,\delta^{3}(\vec{r})
\label{H}
\end{equation}
Each magnetic moment includes an anomalous part. So, hyperfine splitting in the muonium or dimuonium can also shed light on the mystery of the discrepancy between the anomalous magnetic moment of the muon and the theoretical predictions \cite{delaunay_towards_2021}.

\item  Dimuonium is ideal for finding the effects of New Physics \cite{aebischer_effective_2021}. Because from the point of view of effective field theory, New Physics with a scale $\Lambda$ manifests itself in the form of corrections $\mathcal {O}(m^2_\mu /\Lambda^2)$. So, dimuonium has $m^2_\mu/m^2_e \sim $ 44100 times more potential for discovering new physics than positronium or  muonium.

\item It provides an independent test for lepton universality by accurately measuring muon properties in dimuonium experiments. This is very important, because now there are convincing hints of unknown physical effects in this direction. Indeed, for example, a discrepancy arises in rare B-decays.
\begin{equation}
R_{K}=\frac{\operatorname{BR}(B \rightarrow K \mu \mu)}{\operatorname{BR}(B \rightarrow  K e e)}=0.846_{-0.054}^{+0.060}{ }_{-0.014}^{+0.016}
\end{equation}
This LHCb result shows the ratio of B decays in channels with an electron or muon pair in the final state \cite{PhysRevLett.122.191801,aebischer_b-decay_2020}. The first error is statistical, the second is systematic. The deviation of the order of 2.5~$\sigma$ from the theoretical prediction 
$R_ {K}=1$ is quite obvious. 

\item The search for dimuonium is stimulating new experimental methods and accelerator technologies. Indeed, observing and studying dimuonium would be a significant advance, not only as a chance to understand the fundamental mysteries of leptons, but also as a challenge to experimental physicists to develop new discovery tools. 

Studying muon pair production in electron-positron annihilation at the threshold energy will also be useful for estimating the limiting luminosity attainable in a muon collider \cite{Amapane:2019oog}.   

\end{enumerate}

\section{Prerequisites for discovery}
The search for dimuonium is complicated because of large background and small production cross section. The new accelerator concept, called Crab Waist, has proven to be a reliable method of increasing luminosity without excessively increasing colliding beem currents.
It has been successfully tested in DAFNE special runs \cite{zobov_crab_2016} and therefore looks like a promising technique for dimuonium searches.

Budker Institute of Nuclear Physics plans to build a $\mu\mu$-tron facility to observe dimuonium \cite{bogomyagkov_low-energy_nodate}. It is based on a concept similar to Crab Waist with a large crossing angle $\theta\sim 75^\circ$. The large crossing angle allows the created dimuonium to fly away from interaction point before it decays. Let us estimate the decay length $l=c \tau_{\mu\mu} \beta_{\mu\mu} \gamma_{\mu\mu}=$ in this configuration. 

The projections of the initial electron and positron momenta onto flight direction of the dimuonium both equal to  $m_e c\beta_e\gamma_e\sin{\theta}$ (see the Fig.2 in \cite{bogomyagkov_low-energy_nodate} for the definition of the crossing angle $\theta$). 
The created dimuonium moves with the velocity of the center of mass of the original system.
Therefore,
$$\beta_{\mu\mu}=\frac{2(m_e c\beta_e\gamma_e\sin{\theta})c}{2m_e c^2\gamma_e}=\beta_e\sin{\theta}\approx \sin{\theta},$$
since $\beta_e\approx 1$ (the designed beam energy in the $\mu\mu$-tron is 408~MeV). Then $\gamma_{\mu\mu}=1/\cos{\theta}$ and for the dimuonium decay length we get $l=c \tau_{\mu\mu} \tan{\theta}$.

From QED theoretical calculations of dimuonium decay width for a particular state $1^{3} S_{1}$ it is known that $c \tau\approx 543~\mu m$ \cite{Brodsky:2009gx}. For an angle $\theta = 75^\circ$, $\tan{\theta} = 2+\sqrt{3} \approx 3.73$. So, the decay length $l \approx 2~mm$.

Dimuonium search is complicated  by the fact that the background cross-section from the $e^+e^-$ Bhabha scattering is roughly $10^4$ or $10^5$ factor larger than the sought signal from the decay of dimuonium into an electron-positron pair. However, the decay point of dimumium is shifted by 2~mm, and for an estimated vertex resolution of about $350\mu m$ 
this amounts to a displacement of $\frac{2~mm}{350\mu m} \approx 5.7 \sigma$. This circumstance will significantly suppress the Bhabha  background and reduce it to a level sufficient to detect the dimuonium signal. 

The insertion of a thin aluminum foil about 30~$\mu$m thick into the resulting dimuonium beam makes it possible to measure a number of important properties of dimuonium. For example, cross-sections of transitions between 1S and other quantum states of dimuonium, as well as the probability of dimuonium breakup. 

To succeed in interpreting such measurements, we need to know the theoretical QED predictions for these cross sections when dimuonium interacts with the electric field of the target nucleus. The following sections briefly explain the theory used and our new computational algorithm. 

\section{How exotic atom interacts with matter}
The cross sections for elementary atom interactions with an external atomic field in the non-relativistic and Born approximations were calculated in \cite{Mrowczynski:1985qt}.

When dimuonium is scattered in the screened Coulomb field $U(\vec{r})$ of the target nucleus,
a central object through which these cross sections are expressed is the atomic form factor $F_{n_1l_1m_1}^{n_2l_2m_2}$, which is the Fourier transform of $\varphi^*_f(\vec{r}\,)\,\varphi_i(\vec{r}\,)$ with respect to the transferred momentum $\vec{q}$ \cite{dewangan}. Here $\varphi(\vec{r})$ is the Coulomb wave function of the relative motions of muon and antimuon in the dimuonium atom.

In particular, discrete-discrete transition cross section is given by ($V$ is the initial velocity of dimuonium, $\tilde{U}(\vec{q}$)  is the Fourier transform  of $U(\vec{r})$)  \cite{Mrowczynski:1985qt,Alizzi:2021dga}
\begin{equation}
d\sigma_{n_1l_1m_1}^{n_2 l_2 m_2}=\frac{e^2\left(1-(-1)^{l_1-l_2}\right)}
{\pi V^2}\,\left|\tilde{U}(\vec{q})\right|^2\left |F_{n_1l_1m_1}^{n_2 l_2 m_2}
\left(\frac{\vec{q}}{2}\right )\right |^2 qdq,
\label{Trcrs}
\end{equation}
while  the total cross section of dimuonium transitions from the initial $(n,l,m)$ quantum state to some discrete or continuum final states is
\begin{equation}
d\sigma_{nlm}^{tot}=\frac{e^2}{\pi V^2}\,\left|\tilde{U}(\vec{q})\right|^2
\left [1-F_{nlm}^{nlm}(\vec{q}\,)\right ] qdq.
\label{Totcrs}
\end{equation}

A quantum mechanical calculation allows to find the following expression for the atomic form factor \cite{Alizzi:2021dga}:
\begin{equation}
F_{n_{1} l_{1} m_{1}}^{n_{2} l_{2} m_{2}}=N \sum_{l=\left|l_{1}-l_{2}\right|}^{l_{1}+l_{2}} A_{l} I_{l}=N \sum_{s=0}^{\min \left(l_{1}, l_{2}\right)} A_{L-2 s} I_{L-2 s}, 
\label{FF}
\end{equation}
where
\begin{equation}
N=\frac{(2 a)^{l_{1}+1}(2 b)^{l_{2}+1}}{n_{1}+n_{2}} \sqrt{\left(2 l_{1}+1\right)\left(2 l_{2}+1\right) \frac{\left(n_{1}-l_{1}-1\right) !\left(n_{2}-l_{2}-1\right) !}{\left(n_{1}+l_{1}\right) !\left(n_{2}+l_{2}\right) !}}, 
\end{equation}
and
\begin{equation}
A_{l}=i^{l}(-1)^{m_{2}+m} \sqrt{4 \pi(2 l+1)}\left(\begin{array}{ccc}
l_{1} & l_{2} & l \\
0 & 0 & 0
\end{array}\right)\left(\begin{array}{ccc}
l_{1} & l_{2} & l \\
m_{1} & -m_{2} & -m
\end{array}\right) Y_{l m}\left(\Omega_{q}\right). \\
\end{equation}
Here Wigner's $3j$-symbols in the last equation arise from the angular integral of the product of three spherical functions, $L=l_1+l_2$, $s=\frac{1}{2}(L-l)$, while $a$ and $b$ are defined as follows: $$a=\frac{n_2}{n_1+n_2},\;\;b=\frac{n_1}{n_1+n_2}.$$

There are several expressions for the radial integral $I_l$ used in MuMuPy, and the input parameter "Computation method" determines which expression is used to calculate the form factor. If this parameter equals to three, then \cite{dewangan}
\begin{equation}
I_l=\sum\limits_{k=0}^{n_1+n_2-L-2}C_k\,\tilde{J}_k,
\label{IlW}
\end{equation}
where 
\begin{equation}
C_k=\sum\limits_{j=0}^k\frac{(-1)^k(n_1+l_1)!\,(n_2+l_2)!\,(2a)^j(2b)^{k-j}}
{j!\,(k-j)!\,(2l_1+1+j)!\,(2l_2+1+k-j)!\,N_1!\,N_2!}, 
\end{equation}
with $N_1=n_1-l_1-1-j$, $N_2=n_2-l_2-1+j-k$, and 
\begin{equation}
\tilde{J}_k=(2\sigma)^{L-2s}\sum\limits_{p=0}^{s+[\frac{k+1}{2}]}\frac{(-1)^p
(2s+k+1)!\,(L-p+k+1)!\,2^{2(s-p)+k+1}}{[2(s-p)+k+1]!\,p!\,(1+\sigma^2)^{L-p+k+2}}.
\end{equation}
In the last equation,  $[(k+1)/2]$ denotes the integer part of $(k+1)/2$, and $\sigma=\frac{n_1n_2}{n_1+n_2}\,q$. In this case the atomic form factor $F_{n_1l_1m_1}^{n_2l_2m_2}$ is calculated as a four-fold finite series of rational functions of transferred momentum $q$.

The second computation method is based on the results of \cite{Afanasev:1996xz}. In this 
case, the radial integral is given by the expression
\begin{equation}
\begin{gathered}
I_l=\sum\limits_{p=0}^s\sum\limits_{k=0}^{n_1+n+2-L-2}B_{ps}\,H_k\,\left(\frac{2}
{\sigma}\right)^pI_k^{(L,\,p)}(\sigma),
\end{gathered}
\label{IlAT}
\end{equation}
where
\begin{eqnarray} & &
I_{\;k}^{(L,\,p)}(\sigma)=\frac{2\,(2\sigma)^{L-p}\,(L-p+1)!}{(1+\sigma^2)^{L-p+2}}\,
\times \nonumber \\ & &
\left [C_{\;k}^{(L+2,p)}\left(\frac{\sigma^2-1}{\sigma^2+1}\right)+
C_{\;k-1}^{(L+2,p)}\left(\frac{\sigma^2-1}{\sigma^2+1}\right)\right ],
\end{eqnarray}
while
\begin{equation}
B_{ps}=(-1)^{s-p}\,\Gamma(p+1)\binom{s}{p}\binom{L-s+1/2}{p},
\end{equation}
and
\begin{equation}
H_k=C_{n_1-l_1-1,\,n_2-l_2-1}^{2l_1+1,\,2l_2+1}(a,b),
\end{equation}
with
\begin{eqnarray} & &
C_{nm}^{\alpha\beta}(a,b)= \frac{k!\,(n+m-k)!\,}{n!\,m!}\,a^{k-m}\,b^{k-n}\,
\times \nonumber \\ & &
P_{n+m-k}^{(k-m,\,k-n)}(b-a)\,P_{n+m-k}^{(\alpha+k-m,\,\beta+k-n)}(b-a),
\end{eqnarray}
$P_n^{(\alpha,\beta)}(x)$ being Jacobi polynomials. As for the $C_{\;k}^{(\lambda,\,p)}(x)$
functions, they are expressed through the Gegenbauer polynomials:
\begin{equation}
C_{\;k}^{(\lambda,\,p)}(x)=\sum\limits_{l=0}^k\binom{l+2p-1}{l}C_{\;k-l}^{(\lambda-p)}(x).
\label{GGP}
\end{equation}
Finally, when the parameter "Computation method" equals to one, we use our new algorithm, described in the next section, to calculate the radial integral.

\section{Novel computational method for the radial integral}
To calculate the transition form-factor $F_{n_{1} l_{1} m_{1}}^{n_{2} l_{2} m_{2}}$, it is necessary to be able to effectively integrate the radial integral $I_{l}$:
\begin{equation}
I_{l}=\int\limits_0^\infty x^{l_{1}+l_{2}+2} \,e^{-x}\, j_{l}(\sigma x)\, L_{n_{1}-l_{1}-1}^{2 l_{1}+1}(2 a x) \,L_{n_{2}-l_{2}-1}^{2 l_{2}+1}(2 b x) \,d x,
\end{equation}
where $j_l(x)$ is the spherical Bessel Function and $L^m_n(x)$ is the associated  Laguerre polynomial.

In the previous section, we mentioned two methods for calculating this integral. The advantage of the Dewangan's method \cite {dewangan} is its simplicity. However, this leads to the expression (\ref{IlW}), which is less computationally efficient than the expression (\ref{IlAT}), which can be obtained by the much less trivial method of Afanasyev and Tarasov
\cite{Afanasev:1996xz}. The detailed description of both methods can be found in \cite{Alizzi:2021dga}.

In \cite{Alizzi:2021dga}, the radial integral $I_l$ was calculated by a new method that combines the simplicity and straightforwardness of Dewangan's method with the computational efficiency of the Afanasyev and Tarasov method. The results of \cite{alassar_new_2008} on an integral involving the product of Bessel functions and associated Laguerre polynomials were used to develop this novel approach. Here we present a new independent derivation of this new formula for the radial integral $I_l$.

\newtheorem*{thm}{Theorem}
\newtheorem*{lem}{Lemma}
\begin{lem}
Integral of a product, involving polynomial, exponential, and Bessel functions, can be represented as a Hypergeometric function:
\end{lem}
\begin{eqnarray} & &
\int_{0}^{\infty} x^{\gamma-1} e^{-\delta x} J_{\nu}(\mu x) \,d x=\frac{\mu^{\nu}\, \Gamma(\gamma+\nu)}{2^{\nu}\, \delta^{\nu+\gamma} \, \Gamma(\nu+1)}\times \nonumber \\ &&
{ }_{2}F_{1}\left(\frac{\nu+\gamma}{2}, \frac{\nu+\gamma+1}{2}, \nu + 1,-\frac{\mu^{2}}{\delta^{2}}\right)
\label{integlem}
\end{eqnarray}
This integral can be found in \cite{Gradshteyn}, entry 6.621.1. For convenience, below we give an outline of the proof. 
\begin{proof} The idea of the proof is to use several properties of the gamma function, the Bessel function expansion into a series, and the definition of the hypergeometric function in terms of the Pochhammer symbol: 
\begin{enumerate}
    \item Expand Bessel Function $J_{\nu}(x)$ on the left side of equation (\ref{integlem}) 
    in a series
\begin{equation}
J_{\nu}(\mu x)=\sum_{m=0}^{\infty} \frac{(-1)^{m}}{m ! \Gamma(m+\nu+1)}\left(\frac{\mu x}{2}\right)^{2 m+\nu}.
\end{equation}
      \item  Reverse the order of summation and integration. Integration immediately gives by definition of the gamma function the following result:
      \begin{equation}
          \int_{0}^{\infty} x^{ 2 m + \nu + \gamma -1} e^{-\delta x}  d x= \Gamma(2m + \nu + \gamma ) / \delta^{2m + \nu + \gamma}.
      \end{equation}
      Note that term-by-term integration is only justified if $|\mu|<|\sigma|$, because then the final series converges absolutely. However, the end result (\ref{integlem}) has a broader range of applicability by the principle of analytic continuation \cite{Watson}.

     \item Use Legendre's Duplication formula to get rid of $2m$. Since m is a summation index, it requires special attention if we want the final result to be expressed in terms of a hypergeometric function. 
     \begin{equation}
2^{1-2 m-\nu-\gamma} \sqrt{\pi} \,\Gamma(2 m + \nu + \gamma) = \Gamma(m+ \frac{\nu+ \gamma} {2})  \, \Gamma(m + \frac{ \nu + \gamma + 1} {2}).
\end{equation}

     \item Now we have three gamma functions in the combination
     \begin{equation}
         \frac{ \Gamma(m+ \frac{\nu+ \gamma} {2}) \, \Gamma(m + \frac{\nu+ \gamma + 1} 
         {2})}{ \Gamma(m+\nu+1)}.
     \end{equation}
     
     \item Convert each of the three Gamma functions to three Pochhammer symbols $(a)_{m}$ by use of the relation
\begin{equation}
(a)_{m}=\frac{\Gamma(m + a)}{\Gamma(a)}.
\end{equation}
     
     \item Note that exactly three Pochhammer symbols define the hypergeometric function:
\begin{equation}
{ }_{2} F_{1}\left(\frac{v+\gamma}{2}, \frac{v+\gamma+1}{2}, v + 1, z\right) = \sum_{m=0}^{\infty} \frac{(\frac{\nu+ \gamma} {2})_{m}(\frac{\nu+ \gamma + 1} {2})_{m}}{(v + 1)_{m}} \frac{z^{m}}{m !},
\end{equation}
and use again the Legendre Duplication formula to get the final form of the right-hand-side of (\ref{integlem}).

\end{enumerate}
\end{proof}

\begin{thm}
The sought for integral $I_{l}$ is expressed as a two-fold summation involving the quantum numbers and the Hypergeometric function ${ }_{2} F_{1}$:
\end{thm}
\begin{equation}
\begin{gathered}
I_{l}=\frac{2^{l} l !}{(2 l+1) !} \sum_{m_{1}=0}^{n_{1}-l_{1}-1} \sum_{m_{2}=0}^{n_{2}-l_{2}-1}
\frac{(-1)^{m_{1}+m_{2}}(2 a)^{m_{1}}(2 b)^{m_{2}}\left(n_{1}+l_{1}\right) !\left(n_{2}+l_{2}\right) !}
{m_{1} ! m_{2} !\left(n_{1}-l_{1}-1-m_{1}\right) !\left(n_{2}-l_{2}-1-m_{2}\right) !} 
\times\\
\frac{ \left(l+l_{1}+l_{2}+m_{1}+m_{2}+2\right) ! \, \sigma^{l}}
{ \left(2 l_{1}+1+m_{1}\right) !\left(2 l_{2}+1+m_{2}\right) !} 
\times\\
{ }_{2} F_{1}\left(\frac{l+l_{1}+l_{2}+m_{1}+m_{2}+3}{2}, \frac{l+l_{1}+l_{2}+m_{1}+m_{2}+4}{2} ; l+\frac{3}{2} ;-\sigma^{2}\right). 
\end{gathered}
\label{themaineq}
\end{equation}

\begin{proof}
The most obvious method (algorithm) of the proof is the use of the lemma and the expansion of the Laguerre polynomials in a series:
\begin{enumerate}
    \item Express the associated Laguerre Polynomial $L_{n_{2}-l_{2}-1}^{2 l_{2}+1}(2 b x)$ as a series 
\begin{equation}
L_{n_{2}-l_{2}-1}^{2 l_{2}+1}(2 b x)=\left(n_{2}+l_{2}\right) ! \sum_{k=0}^{n_{2}-l_{2}-1} \frac{(-1)^{k}(2 b x)^{k}}{k !\left(n_{2}-l_{2}-1-k\right) !\left(2 l_{2}+1+k\right) !}.
\end{equation}

    \item In the same way, express the second associated Laguerre Polynomial $L_{n_{1}-l_{1}-1}^{2 l_{1}+1}(2 a x)$ as a finite series.
    
    \item Use the Lemma to integrate the product of the power, exponential and Bessel functions.
    
    \item Use
    \begin{equation}
    \Gamma\left(l+\frac{3}{2}\right )=\frac{(2l+1)!\,\sqrt{\pi}}{2^{2l+1}\,l!},    
    \end{equation}
    and get the final expression (\ref{themaineq}) for $I_{l}$.
    
    \end{enumerate}
\end{proof}

\section{The Fortran core of MuMuPy}
Fortran is a pioneer of symbolic languages: The first Fortran compiler was developed in 1957 by John Backus and his team at IBM in San Jose, California \cite{Perkel}. Despite its solid age, Fortran is still widely used in climate modelling, computational chemistry, hydrodynamics and other areas of scientific computing  \cite{Perkel,ott_fortran-keras_2020}.
The reason why Fortran remains one of the most widely used language in scientific computing is its high performance in numerical computations and a huge amount of ready and well-tested scientific Fortran code in the world \cite{Why_Fortran}.

In our case, no complicated programming is required to code formulas such as (\ref{Trcrs}) and (\ref{Totcrs}), and Fortran is a natural choice for efficient numerical calculations based on these formulas. 

The two main Fortran subunits of the MuMuPy code are the two double precision functions TCRS(Z,n,l,m) and TRCRS(Z,n1,l1,m1,n2,l2,m2), which calculate the total and the transition cross sections, respectively. Other Fortran subunits provide all the supporting computations needed. For example, double precision function U(q) calculates the Fourier transform of the atomic potential in the Thomas-Fermi-Moli\'{e}re approximation \cite{Mrowczynski:1985qt}:
\begin{eqnarray} & &
U(q)=4\pi Z e\sum_{i=1}^3\frac{\alpha_i}{q^2
+\beta_i^2},\;\;\;\beta_i=\frac{m_e b_i}{121}Z^{1/3}. \nonumber \\ &&
b_1=6.0,\;\;b_2=1.2,\;\;b_3=0.3,\;\;
\alpha_1=0.10,\;\;\alpha_2=0.55,\;\;\alpha_3=0.35. \qquad
\end{eqnarray}
The generalized Gegenbauer polynomials $C_k^{(\lambda,\,p)}(x)$ in (\ref{IlAT}) are efficiently calculated using the recurrence relation \cite{Afanasev:1996xz,Alizzi:2021dga}
\begin{eqnarray} & &
(k+2)C_{k+2}^{(\lambda,\,p)}(x)=\left[k+1+2p+2x(k+\lambda-p+1)\right]C_{k+1}^{(\lambda,\,p)}
(x)-\nonumber \\ & & \left[k+2\lambda-2p+2x(k+\lambda+p)\right ] C_k^{(\lambda,\,p)}(x)+(k+2\lambda-1)C_{k-1}^{(\lambda,\,p)}(x), \qquad 
\end{eqnarray}
with the initial values 
\begin{eqnarray} & &
C_{\;0}^{(\lambda,\,p)}(x)=1,\;\;C_{\;1}^{(\lambda,\,p)}(x)=2\left[p+(\lambda-p)x\right],
\nonumber \\ &  &
C_{\;2}^{(\lambda,\,p)}(x)=2(\lambda-p)(1+\lambda-p)x^2+4p(\lambda-p)x+
2p(1+p)-\lambda. \qquad
\end{eqnarray}
The hypergeometric function in (\ref{themaineq})  can be expressed in terms of Jacobi polynomials \cite{Alizzi:2021dga}:
\begin{eqnarray} && 
{_2F_1}\left(\frac{N_1+1}{2},\,\frac{N_1+2}{2};\,
l+\frac{3}{2};\,-\sigma^2\right)= \nonumber \\ &&
\left \{\begin{array}{l} \left (\cos{\phi}\right )^{2(l+M+2)}\;
\frac{P_{\;M}^{\left (l+\frac{1}{2},\,\frac{1}{2}\right )}(\cos{2\phi})}
{P_{\;M}^{\left (l+\frac{1}{2},\,\frac{1}{2}\right )}(1)},\;\; \mathrm{if}\;\;
N_1-2(l+1)=2M, \\ \\ \left (\cos{\phi}\right )^{2(l+M+2)}\;
\frac{P_{\;M+1}^{\left (l+\frac{1}{2},\,-\frac{1}{2}\right )}(\cos{2\phi})}
{P_{\;M+1}^{\left (l+\frac{1}{2},\,-\frac{1}{2}\right )}(1)},\;\; \mathrm{if}\;\;
N_1-2(l+1)=2M+1,\end{array} \right . \qquad
\end{eqnarray}
where the angle $\phi$ is defined by $\tan{\phi}=\sigma$ and $M$ is an integer. The Jacobi polynomials themselves are efficiently calculated using the three-term recurrence relation 
\begin{eqnarray} &&
P_{\;n+1}^{(\alpha,\,\beta)}(x)=
\left (\frac{(2n+\alpha+\beta+1)(2n+\alpha+\beta+2)}
{2(n+1)(n+\alpha+\beta+1)}\,x+\right . \nonumber \\ && \qquad \qquad \qquad \left .
\frac{(\alpha^2-\beta^2)(2n+\alpha+\beta+1)}
{2(n+1)(n+\alpha+\beta+1)(2n+\alpha+\beta)}\right )P_{\;n}^{(\alpha,\,\beta)}(x)-
\nonumber \\ && \qquad \qquad \qquad \quad
\frac{(n+\alpha)(n+\beta)(2n+\alpha+\beta+2)}{(n+1)(n+\alpha+\beta+1)
(2n+\alpha+\beta)}\,P_{\;n-1}^{(\alpha,\,\beta)}(x), \qquad \qquad 
\end{eqnarray}
with the following initial values:
\begin{equation}
P_{\;0}^{(\alpha,\,\beta)}(x)=1,\;\;\; P_{\;1}^{(\alpha,\,\beta)}(x)=\frac{1}{2}\left [
(\alpha+\beta+2)x+\alpha-\beta\right].
\end{equation}
In MuMuPy we use two programs from the CERNLIB library: DGAUSS and DWIG3J. The First, DGAUSS(F,a,b,eps) allows to integrate a user-defined function F from a to b with a given precision eps using the adaptive Gaussian quadrature method. The second, DWIG3J(j1,j2,j3,m1,m2,m3), computes the Wigner 3j symbol: 
\begin{equation}
\left(\begin{array}{ccc}
j_{1} & j_{2} & j_{3} \\
m_{1} & m_{2} & m_{3}
\end{array}\right) \equiv \frac{(-1)^{j_{1}-j_{2}-m_{3}}}{\sqrt{2 j_{3}+1}}\left\langle j_{1} m_{1} j_{2} m_{2} \mid j_{3}\left(-m_{3}\right) j_{1} j_{2}\right\rangle.
\end{equation}
We have embedded slightly modified versions of them in the CrossSection.f code, so that the user does not need the CERNLIB library to compile it, for example, for further use in Python: \begin{lstlisting}
  $ gfortran -shared  CrossSection.f -o CrossSection.so -fPIC
\end{lstlisting}

\section{Fortran bindings for Python}
Python is an increasingly popular programming language among scientists \cite{Python}. The reasons of its popularity are that Python is easier to learn for beginners, and also its versatility thanks to its many packages such as NumPy and SciPy. 

However, due to the dynamic nature of Python, it is much slower in computational tasks like ours than compiled languages like C ++ or Fortran. It is expected that in future a new programming language Julia, which combines the best features of scripting and compiled languages, will become a golden standard in scientific computing and data analysis \cite{Julia,Julia1,Julia2}. But Julia is currently less versatile and less documented than Python.

In MuMuPy we use Fortran for computations and Python for GUI. There are several ways to combine the best of these two languages. To do this, we use the built-in Python library called ctypes. Ctypes is ``a foreign functions library" for Python and it makes it easy to import dynamic libraries, call functions, access C-style data types. Note the table of relations Tab.\ref{ctypestab} between the corresponding C-style data types and Python data types. 
\begin{table}[ht]
\centering
\begin{tabular}{|l|l|l|}
\hline ctypes type & C type & Python type \\
\hline c\_bool & \_Bool & bool (1) \\
\hline c\_char & char & 1-character bytes object \\
\hline c\_wchar & wchar\_t & 1-character string \\
\hline c\_wchar\_p & wchar\_t * (NUL terminated) & string \\
\hline c\_double & double & float \\
\hline c\_float & float & float \\
\hline c\_int & int & int \\
\hline
\end{tabular}
\caption{Some fundamental Data Types in ctypes.}
\label{ctypestab}
\end{table}
As was already mentioned, computational strategy outlined in the theoretical part of this article resides in the Fortran module.  All three computational methods were implemented independently of each other to thoroughly check the correctness of the code. We created a Python wrapper for our Fortran code so that it can be easily used alongside common data processing routines: graphing, machine learning, interactive computing. So first we compile the Fortran module into a shared library using gfortran. Then, using ctypes in Python, we load the library into our Cross-Section calculator class, MuMuPy (see Fig.\ref {classmumu}). 
\begin{figure}[ht]
\begin{lstlisting}[language=Python]
from ctypes import cdll
from ctypes import *

class MuMuPy:
    def __init__(self, pathToBinary):
        self.pathToBinary = pathToBinary
        redirect_stdout()        
        ###IMPORTING THE COMPILED FORTRAN LIBRARY
        try:
            self.libc = cdll.LoadLibrary(pathToBinary)
        except:
            print("Library not found. Please specify .so") 
\end{lstlisting}
\caption{Importing a shared library via ctypes in MuMuPy class.}
\label{classmumu}
\end{figure}

Note that passing arguments by reference allows functions in the shared library to modify their contents. This idea makes it easy to return multiple output variables from a Fortran function. 

\section{A cloud experience}
To assess which cloud interface is most suitable for computing particle physics, we looked at a very close area - data science. The main questions we asked ourselves were whether a particular cloud service was used in research and which cloud services are best suited to support the underlying goals of physical computing or data analysis. For the computational physicist, turning computing into an interactive cloud application achieves several goals: 
\begin{enumerate}
    \item Popularization of research
    \item Cross-platform compatibility
    \item Ease of testing
    \item Visual verification of results
    \item Discovering higher-level patterns in data
\end{enumerate}
We will briefly review the most popular solutions, and then demonstrate how we built the MuMuPy interactive dimuonium Cross-Section calculator.
\begin{table}[ht]
\centering
\begin{tabularx}{\textwidth}{ |l|X|l| }

\hline Framework & Summary  \\
\hline \href{https://docs.streamlit.io/}{Streamlit} & Turns data scripts into shareable web apps in minutes. The best option for rapid-prototyping. Compact and transparent code makes it intuitive to get started as soon as possilbe.\\
\hline \href{https://dash.plotly.com/}{Dash} & Low-code framework for building production-ready data apps for a bigger organization   \\
\hline \href{https://flask.palletsprojects.com/ }{Flask} & Web framework to handle the requests. Very customizable, but need building UI from pure HTML and Js.  \\
\hline \href{https://github.com/voila-dashboards/voila}{Voila} & Jupyter notebooks become standalone web applications with Voil\'{a}.
Unlike traditional HTML-to-Jupyter notebooks, each user that connects to the Voil\'{a} tornado application receives their own Jupyter kernel.  \\

\hline \href{https://gradio.app/}{Gradio} & With only a few lines of code, one may create an easy-to-use demo for a machine learning model or function. 
Gradio lets quickly build customisable UI components for TensorFlow or PyTorch models, as well as arbitrary Python functions.  \\
\hline \href{https://panel.holoviz.org/}{Panel} & Panel - a Python toolkit that allows to build bespoke interactive web apps and dashboards by connecting user-defined widgets to plots, pictures, tables, and text.

Panel is unique among other alternatives in that it supports practically all charting libraries. Allows multi-page applications easily.  \\
\hline
\end{tabularx}%
\caption{A summary of dashboarding frameworks in Python.}
\label{frameworkta}
\end{table}

In general, the rivals are equally good. However, each choice of framework reflects slightly different niche needs. See Table Tab.\Ref{frameworkta} for a complete comparison. For the cross section calculator, after testing several alternatives, we chose Streamlit. This framework allowed us to turn computation into an application and iterate as quickly as possible. With faster prototyping, we could focus more on physics without wasting precious time. So, with Streamlit, creating an interface and reacting to changes is as easy as declaring integer variables equal to the slider components. 

This simple variable assignment (see Fig.\ref{streamlitsliders}) not only  displays the slider on the webpage, but when the user interacts, re-runs dependent sections of the code.  There is no need to write an event loop. There is no need to define the control flow.  Streamlit automatically handles component states as well as UI logic and implementation.
\begin{figure}[ht]
\begin{lstlisting}[language=Python]
import streamlit as st
Zlit = st.slider('Atomic number Z', 1, 137, 1)
nlit = st.slider('Main quantum number n', 1, 10, 1) 
llit = st.slider('Angular quantum number l', 0, 10, 0) 
mlit = st.slider('Angular projection m', 0, 10, 0) 
#It means int Zlit equals a value from the slider ranging from 1 to 137 with default value 1 
\end{lstlisting}
\caption{Adding and handling sliders to the MuMuPy calculator web interface.}
\label{streamlitsliders}
\end{figure}

As a consequence, any component change triggers a restart, so all variables are reinitialized. It is desirable to have a cumulative table (or a Pandas DataFrame). In addition, it is desirable  to store all resulting cross sections for each parameter combination $(Z, n, l, m)$. For this we use built-in streamlit cache decorator (see Fig.\ref{cachedecorator}).  Caching speeds up computationally-expensive functions by hashing and storing results, and retrieving results on demand instead of re-evaluating them. Please note that the storage takes place by reference.
\begin{figure}[ht]
\begin{lstlisting}[language=Python]
@st.cache(allow_output_mutation=True)
def get_data():
    return []
    
\end{lstlisting}
\caption{Streamlit caching decorator for persistent variable storage.}
\label{cachedecorator}
\end{figure}

Unexpectedly, we can benefit from this caching behavior in a different way. At first, adding the results to a function seems pointless. In fact, this procedure makes more sense than is seen on the surface, because get\_data calls are cached. As a result, the code shown in Fig.\ref {buttoncode} means that when the button ``Add to table'' is clicked, we change the list cached by reference. So the list is growing and going through repeated runs. Finally, when clicking on the download the spreadsheet button, the researcher receives an editable, ready-to-use table.
\begin{figure}[ht]
\begin{lstlisting}[language=Python]
if st.button('Add to table'):
    get_data().append({'Z': Zlit, 'n': nlit, 'l':llit, 'm':mlit, 'crs': res.value})
if st.button('Clear last'):
    get_data().pop()
if st.button('Clear all'):
    get_data().clear()
\end{lstlisting}

\caption{Accumulation of the results of cross-section calculations.}
\label{buttoncode}
\end{figure}

\section{Conclusion}
In this article, we introduced MuMuPy, a handy Python package for calculating cross sections for dimuonium-matter interactions in the non-relativistic approximation. The package was developed for future dimuonium research planned at our home institute. However, we hope that other researchers around the world who are interested in exploring true muonium can also benefit from it. For this purpose, we have developed a cloud-based tool to make MuMuPy easier to use.

Three alternative methods for calculating the atomic form factor have been implemented as Fortran code in MuMuPy. Since they all give the same results, this reinforces our belief in the reliability of MuMuPy. 

We use dimuonium atomic units throughout the paper, where $c=\hbar=1$, the unit of mass is $\frac{1}{2}m_\mu$(reduced mass in the dimuonium atom), and the unit of length is the radius of the first Bohr orbit in dimuonium.

Note that for $ n> 13 $, as a rule, MuMuPy cannot give a reliable result due to the cancellation of large numbers in alternative series representing the atomic form factor. This is a common problem in all known implementations of atomic form factors \cite{Silva},  and Rydberg states require special methods to cope with it \cite{dewangan}.

\section{Grants and Acknowledgements}
The work is supported by the Ministry of Education and Science of the Russian
Federation and in part by RFBR grant 20-02-00697-a.

\bibliographystyle{elsarticle-num}
\bibliography{library}

\end{document}